\documentclass[prb,twocolumn,superscriptaddress,amsmath,amssymb]{revtex4}
\newcommand{\fatr}{\mathbf{r}}

\usepackage{array}
\usepackage{graphicx}
\usepackage{color}
\usepackage{float}
\usepackage{threeparttable}
\usepackage{amsmath}
\usepackage{algorithm}
\usepackage[noend]{algpseudocode}
\usepackage{array}
\newcolumntype{L}[1]{>{\raggedright\let\newline\\\arraybackslash\hspace{0pt}}m{#1}}
\newcolumntype{C}[1]{>{\centering\let\newline\\\arraybackslash\hspace{0pt}}m{#1}}
\newcolumntype{R}[1]{>{\raggedleft\let\newline\\\arraybackslash\hspace{0pt}}m{#1}}

\def\beq{\begin{equation}}
\def\eeq{\end{equation}}
\def\bea{\begin{eqnarray}}
\def\eea{\end{eqnarray}}
\def\fatR{{\bf R}}
\def\fatr{{\bf r}}

\def\fatk{{\bf k}}

\begin{document}
\title{
Al$_x$Ga$_{1-x}$As crystals with direct 2\:eV band gaps  from computational alchemy
}
\author{K. Y. Samuel Chang}
\affiliation{Institute of Physical Chemistry and National Center for Computational Design and Discovery of Novel Materials (MARVEL), Department of Chemistry, University of Basel, 4056 Basel, Switzerland}
\author{O. Anatole von Lilienfeld}
\email{anatole.vonlilienfeld@unibas.ch}
\affiliation{Institute of Physical Chemistry and National Center for Computational Design and Discovery of Novel Materials (MARVEL), Department of Chemistry, University of Basel, 4056 Basel, Switzerland}

\date{\today}

\begin{abstract}
\noindent
{\color{black} We use alchemical first order derivatives for the rapid yet robust prediction of band structures.
The power of the approach is demonstrated for the design challenge of finding} Al$_x$Ga$_{1-x}$As {\color{black} semiconductor alloys}
with large direct band gap using computational alchemy within a genetic algorithm.
Dozens of crystal polymorphs are identified for $x>\frac{2}{3}$ with direct band gaps larger than 2\:eV according to HSE approximated density functional theory. 
Based on a single generalized gradient approximated density functional theory band structure calculation of pure GaAs we observe convergence after visiting only $\sim$800 
crystal candidates. 
The general applicability of alchemical gradients is demonstrated for band structure estimates in III-V and IV-IV crystals as well as for
H$_2$ uptake in Sr and Ca-alanate crystals.
\end{abstract}

\maketitle

\section{Introduction}
\begin{figure}
\centering
\includegraphics[width=8.5cm]{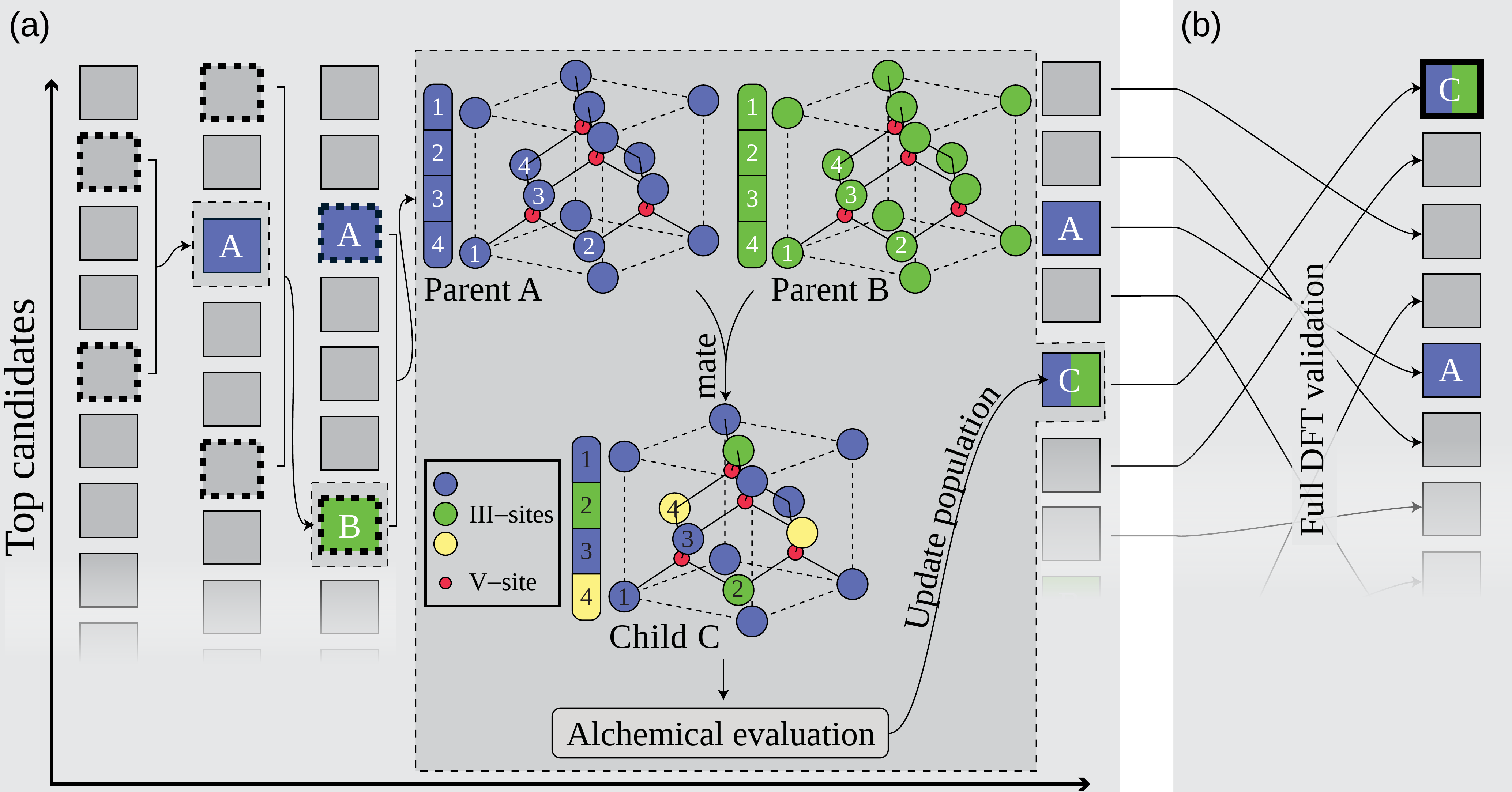}
\caption{
Illustration of hybrid alchemical perturbation based genetic optimization used for Al$_x$Ga$_{1-x}$As alloys in 3x3x3 and 4x4x4 supercells.
(a) Given 20 top crystal candidates (parents) with largest direct band gap, children are iteratively bred until they outperform at least one of the 20 top parents.
The inset exemplifies the mating of randomly picked parent A and B which produce a top-performing child C in the third generation. 
C subsequently enters the updated pool of parents.
This process is repeated until convergence, based on rankings exclusively obtained through alchemical perturbation estimates. 
Initialization corresponds to a single unperturbed self-consistent DFT calculation and 20 random perturbations. 
(b) Self-consistent DFT validation is carried out for all converged parents, leading to a slightly altered ranking.
}
\label{fig:genetic_algorithm}
\end{figure}

Major advancements of many technologies hinge upon the performance of underlying materials. 
The efficiency of photovoltaic materials, for example, heavily depends on the electronic properties of the solid-state or organic semiconductors, 
or the stability of energy storage materials is governed by their energy landscapes and vibrational modes.
With the help of modern compute hardware and {\it ab initio} theories, properties of materials can nowadays be predicted {\it in silico} before their assessment is carried out experimentally.
\cite{Capasso_science_1987,
Beratan_science_1991, 
Norskov_prl_2002,
Jorgensen_science_2004,
Gunsteren_pnas_2005,
Rafael_2015,
ECarter_jmca_2013}
Subsequent synthesis and characterization is costly and should focus only on those materials predicted to exhibit the most promising properties.
Computational materials design procedures therefore hold much promise to lift some of the chemical challenges the world is facing today. 

Finding materials with superior properties in chemical compound space (CCS), the space consisting of all possible materials,\cite{CCS_nature_2004, Anatole_ijqc_2013} can be seen as a numerical optimization problem.
Due to the immense amount of possible compounds and atomic configurations in CCS naive screening is prohibitive.
The combinatorial nature of CCS implies that fast yet accurate property estimates, as well as efficient algorithms for searching, are 
desirable.\cite{Zunger_1999,
WYang_jacs_2006,
Reiher_ijqc_2014,
Best_first_search_Geerlings_2016}
Alchemical derivatives provide useful, chemically local, gradient information, suggesting the applicability of gradient-based algorithms to the design new materials.
Alchemical derivatives were already employed to model the stability of binary solid mixtures within virtual crystal approximations~\cite{Marzari_prl_1994,Zunger_prb_2008}. 
They have also afforded accurate predictions of various properties and compound classes within less approximate perturbation theory~\cite{Anatole_prl_2005, Anatole_jcp_2006, Anatole_jcp_2009, Geerlings_jctc_2013, Geerlings_csr_2014, Yasmine_jcp_2017, Ayers_ijqc_2017, Karth_jpcl_2017}.
{\color{black} Here, we present numerical evidence which suggests that alchemical first order derivatives can also be used
for the rapid yet robust prediction of band structures.}
For molecules and ionic crystals, even chemical accuracy can be achieved in terms of alchemical first order based estimates of relative energies
when fixing number of valence electrons and geometry.\cite{JCP_2016, Alisa_2016}
Such constraints can also be met by several material classes of great interest, such as III-V semiconductors.
Here, we rely on a materials design algorithm which combines alchemical gradients with stochastic sampling and which holds promise for 
general computational materials design campaigns.
Using this algorithm, we have optimized III-V solid {\color{black} alloy solution} mixtures consisting of Al$_{x}$Ga$_{1-x}$As with respect to band structure, 
arguably one of the most important properties of semiconductors due to their many electronic applications.\cite{band_structure_1, Anatole_prb_2008, band_structure_2, band_structure_3} 
The algorithm is illustrated in Fig.~\ref{fig:genetic_algorithm}. 

\section{Theory}
\begin{figure}
\centering
\includegraphics[width=8.5cm]{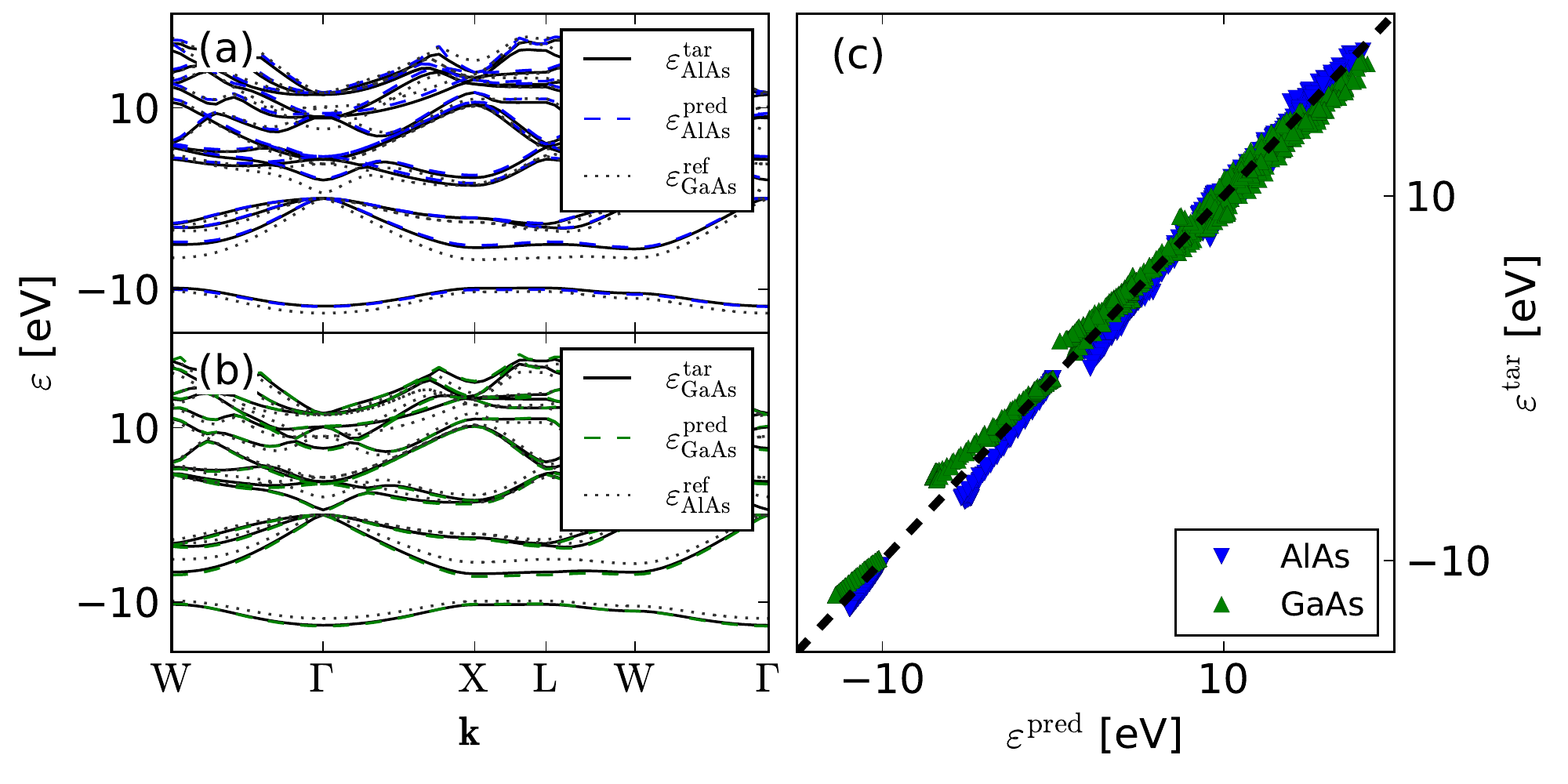}
\caption{
AlAs Band structures (solid) predicted from alchemical predictions (dashed) using GaAs (dotted) as reference (a). 
GaAs Band structures (solid) predicted from alchemical predictions (dashed) using AlAs (dotted) as reference (b). 
Fermi level shifted to zero.
(c) Scatter plot of alchemical prediction on every $\fatk$-point for all unoccupied and occupied bands.
The mean absolute error for the latter amounts to 0.11 eV.
}
\label{fig:band_structure}
\end{figure}

Alchemical gradients are obtained as first order perturbation of the ground state energies and resulting from a linear interpolation between reference (ref) Hamiltonian, $\hat{H}^{\rm ref}$, and target (tar) Hamiltonian, $\hat{H}^{\rm tar}$, 
\beq
E^{\rm tar}\approx E^{\rm pred} = \langle\Psi^{\rm ref}|\hat{H}^{\rm tar}|\Psi^{\rm ref}\rangle
\eeq
where $|\Psi^{\rm ref}\rangle$ is the wavefunction of the reference system.\cite{Anatole_jcp_2009,Anatole_ijqc_2013,JCP_2016}
To estimate changes in band structure, the same formula is applied to each eigenvalue of $\hat{H}^{\rm tar}$ at any given wavevector $\fatk$ using the corresponding eigenfunction $|\phi_{\fatk}^{\rm ref}\rangle$.
In Fig.~\ref{fig:band_structure}, the predictive performance for band structures is illustrated using GaAs (AlAs) as a reference in order to predict AlAs (GaAs):
Note how most of the details in the band structure are faithfully reproduced, and how the estimates can even account for the change from direct band gap 
with conduction band minimum at $\Gamma$ for GaAs to indirect band gap with minimum at $\rm X$ for AlAs. 
The prediction error $\Delta\varepsilon_{\rm LUMO}(\Gamma)=\varepsilon^{\rm pred}_{\rm LUMO}(\Gamma) - \varepsilon^{\rm tar}_{\rm LUMO}(\Gamma)$ amounts to 0.25 (-0.2) eV,
and $\Delta\varepsilon_{\rm LUMO}({\rm X}) =$ 0.14 (-0.15) eV for AlAs (GaAs) using GaAs (AlAs) as a reference.
This implies a low band gap prediction error of only $\Delta E_g =$ 0.14 eV and $\Delta E_g =$-0.2 eV for AlAs and GaAs, respectively.
A scatter plot for estimates corresponding to every $\fatk$-point and every band is shown in Fig.~\ref{fig:band_structure}(c), indicating a very good correlation.
The mean absolute error (MAE) made when predicting eigenvalues for all occupied bands and $\fatk$-points is as low as 0.11 eV.

\begin{figure}
\centering
\includegraphics[width=8.5cm]{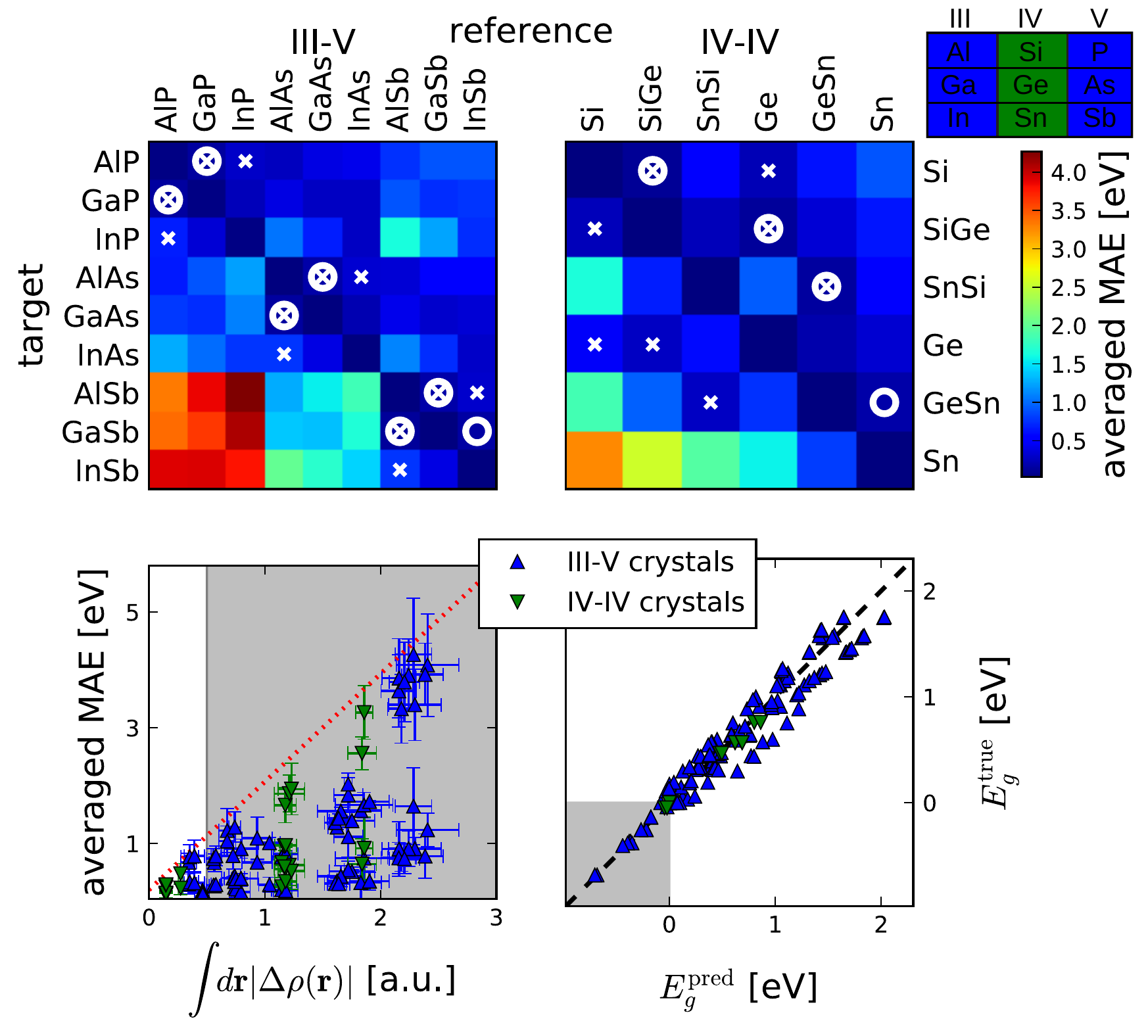}
\caption{
Averaged MAEs for alchemical band structure predictions among III-V (upper left) and IV-IV (upper right) semiconductors.
Errors less than 0.2 eV 
and combinations with ${\int d\fatr |\Delta\rho(\fatr)|<0.5\:\mbox{a.u.}}$ are highlighted by white circles and crosses, respectively.
Bottom left: Averaged MAE vs.~integrated density difference for III-V and IV-IV semiconductors.
The error bars correspond to variations due to changes in lattice constants.
Bottom right: Predicted vs.~true band gap (gray area corresponds to ${\int d\fatr |\Delta\rho(\fatr)|<0.5}\:\mbox{a.u.}$).
}
\label{fig:heatmap}
\end{figure}

Such predictive accuracy is no coincidence. Averaged MAEs for alchemical band structure predictions 
among all possible III-V and IV-IV semiconductors are provided in Fig.~\ref{fig:heatmap}, 
Not surprisingly, predictions ``close-by'', e.g.~among III elements for the same V element, are generally very accurate,  
and largest errors are found for combinations associated with dramatic
changes in electronic states, e.g.~estimates of III-Sb or Sn crystals using III-P or Si as respective reference.
Integrated changes in electron density, shown in Fig.~\ref{fig:heatmap}, are found to correspond to an upper error bound,
suggesting the possibility to construct meaningful error measures by means of inexpensive approximate estimates of 
electron density changes. 

\section{Computational details}
\label{sec:detail}
Alchemical derivative and PBE results are computed using the {\tt ABINIT} package\cite{ABINIT} with Goedecker norm-conserving pseudopotentials\cite{Krack_2005, Goedecker_1998} and planewave cutoff of 100 Ry.
Monkhorst-Park\cite{MonkhorstPack} ${6\times 6\times 6}$ mesh for fcc primitive cell band structure, and ${3\times 3\times 3}$ mesh for band structure optimization (both for ${3\times 3\times 3}$ and ${4\times 4\times 4}$ fcc supercell) are used respectively.
{\tt VASP}\cite{VASP} package with PAW\cite{PAW} pseudopotentials with default cutoff are used for HSE\cite{HSE} results for Al$_x$Ga$_{1-x}$As ${3\times 3\times 3}$ fcc supercell.
10 and 16 virtual orbitals are used for fcc primitive and supercells respectively.
Experimental lattice constants (if available) are used for every crystal considered in this report.
Otherwise calculated values are used from the literature (see table~\ref{tab:a_list}).
Alchemical derivatives are printed by restarting with the reference wavefunction in
{\tt ABINIT}, and setting the SCF iteration step to 0 to evaluate $\partial_\lambda E$.
Three independent optimizations have been carried out for $3\times 3\times 3$ fcc supercells,
and one optimization for a $4\times 4\times 4$ fcc supercell.
Gradient sampling is carried out with population size of 20 and mutation rate of 0.5\%.
Pseudocode and convergence criteria have been included in the appendix.
The $3\times 3\times 3$ fcc supercell crystal is extended to $9\times 9\times 9$ fcc supercell to compute the periodic Coulomb matrix in order to account for all possible periodic images.

\section{Results and discussion}
\subsection{Band structures of III-V and IV-IV Semiconductors}

Alchemical prediction from the reference material of lattice constant $a$ to any target material (tar) on the $i^{th}$ band at a given crystal momentum $\fatk$ is constructed as follows:
${
\varepsilon_{i, a}^{\rm pred}(\fatk) = \varepsilon_{i, a}^{\rm ref}(\fatk) + \partial_\lambda\varepsilon_{i, a}^{{\rm ref}\rightarrow{\rm tar}}(\fatk),
}$
where the derivative is a 3D integral, ${\partial_\lambda\varepsilon_{i, a}^{{\rm ref}\rightarrow{\rm tar}}(\fatk) = \langle\phi_{i\fatk}^{\rm ref}|H_{\rm T} - H_{\rm R}|\phi_{i\fatk}^{\rm ref}\rangle_{a}}$, calculated from the $i^{th}$ orbital of the reference material of lattice constant $a$ with crystal momentum $\fatk$.
The prediction of band gap is defined as:
\beq
E_{g, a}^{\rm pred} = \displaystyle\min_{\fatk}\big(\varepsilon_{{\rm LUMO},a}^{\rm pred}(\fatk) - \varepsilon_{{\rm HOMO},a}^{\rm pred}(\Gamma)\big),
\eeq
where the maximum of conduction band of materials investigated is located at $\Gamma$-point.
A predicted band gap is direct if the predicted minimum of $\varepsilon_{\rm LUMO}$ is located at $\Gamma$-point.
Lattice constants for the semiconductors are scanned from 5.4 \AA\ to 6.5 \AA\ in step of 0.05\:\AA\ including extra points correspond to the lattice constant reported in the literature. 
This range covers all equilibrium constants reported for all the semiconductors considered.

We quantify the performance of our predictions using the mean absolute error (MAE) of a prediction to the entire $i^{th}$ band of a target material at lattice constant $a$, made from some reference material as
\beq
\label{eq:error_mae}
\begin{array}{lcl}
{\rm MAE}_{i,a}({\rm ref, tar}) &=& \displaystyle\sum_{\mathbf{k}}|\varepsilon_{i, a}^{\rm pred}(\mathbf{k})-\varepsilon_{i, a}^{\rm true}(\mathbf{k})|w(\mathbf{k}),
\end{array}
\eeq
where $\varepsilon_{i,a}(\mathbf{k})$ is the $i^{th}$ eigenvalue of Hamiltonian at lattice constant $a$ and crystal momentum $\mathbf{k}$.
$w(\mathbf{k})$ is the corresponding Monkhorst-Pack\cite{MonkhorstPack} weight for the sampled special k-points.
Throughout this paper, the error reported corresponds to average over all bands and lattice constants.
The subscripts $i$ and $a$ are omitted unless noted otherwise.

The averaged MAE over all lattice constants and bands for all combinations of reference/target pair, or alchemical path, is on display in the upper panels in Fig.~\ref{fig:heatmap}.
Overall, decent agreements are found with most predictions being in agreement with the target by less than 0.5 eV.
Alchemical estimates with averaged MAE less than 0.2 eV are highlighted by the white circles.
A general trend can be observed:
Predictions using semiconductors containing $3^{rd}$-row elements in the reference give overall the largest averaged MAE, when the target material has elements from the $5^{th}$ row, as can be seen by the red/orange corner of the upper panels (left for III-V semiconductors and right for IV-IV semiconductors) in Fig.~\ref{fig:heatmap}.
A similar target-reference pattern has been observed for alchemical predictions of covalent bonding, and is due to lack of similarity of electron densities.\cite{JCP_2016, Alisa_2016}

A scatter plot between averaged MAE and integrated absolute electron density difference, defined as ${|\Delta\rho(\fatr)|=|\rho_{\rm tar}(\fatr) - \rho_{\rm ref}(\fatr)|}$, is shown in the lower left panel in Fig.~\ref{fig:heatmap}.
The results suggest that there is an upper error bound:
Any alchemical path with small density changes will give a small predictive error to band structure, as all the points lie beneath the red dotted line.
However, a small error does not necessarily imply small density changes, as many results with small predictive error correspond to large density changes.
This is because the band structure is determined by the relative differences in Hamiltonian eigenvalues and the corresponding orbital structure.
Cancellation of higher order (curvature) effects along the alchemical path can lead to a small predictive error for first order estimates yet large density change.
Similar error cancellation has also been discussed in the context of covalent bonding Ref.~\onlinecite{JCP_2016}.
The fact that small density changes imply accurate predictions can be used as a sufficient condition to detect good predictive power.
This might be useful for future studies if decent approximations to inexpensively estimate electron density changes can be found.
The region with $\int d\fatr|\Delta\rho(\fatr)| < 0.5\:\mbox{a.u.}$ electron/primitive cell is highlighted by the white background in the lower left panel.
The band gap predictions of the corresponding alchemical paths highlighted as white crosses in the upper panels of Fig.~\ref{fig:heatmap}.
A scatter plot versus the true band gap is shown in the lower right panel,
remarkable agreement is found with MAE of less than 0.05 eV.
The linear fit gives $E^{\rm true}_g \approx 0.93E^{\rm pred} + 0.019\:$eV, with $R^2=0.93$, $\mbox{RMSE}=0.047$ eV and $\mbox{MAE}=0.036$ eV.
The negative $E_g$ in the lower right panel of Fig.~\ref{fig:heatmap} correspond to GaSb and InSb with small lattice constants.
In such environment, the conduction band minimum is lower than the valence band maximum.
Notice that decent predictive power can be achieved by alchemical predictions, satisfying the sufficient condition above, even at such extreme situation.

It is possible to use III-V semiconductor reference calculations to predict the band structure of other IV-IV semiconductors.
However, such interpolations do not provide satisfactory predictive power when using only first order alchemical derivatives, due to the dissimilarity of the electronic density. Similar conclusions hold when attempting to predict band structures of II-VI semiconductors using III-V reference densities.

\begin{table}
\centering
\caption{Lattice constants of III-V and IV-IV semiconductors in \AA. Most of the lattice constants are taken from Ref.~\onlinecite{semiconductor_parameters}, while SnSi from Ref.~\onlinecite{SnSi}, GeSn from Ref.~\onlinecite{GeSn}, and SiGe from Ref.~\onlinecite{SiGe}}
\begin{tabular}{cc|cc|cc||cc|cc}
 \hline
 \hline
  AlP & 5.464 & AlAs & 5.660 & AlSb & 6.136 &   Si & 5.431 & SiGe & 5.432 \\
 \hline
  GaP & 5.451 & GaAs & 5.654 & GaSb & 6.090 &   Ge & 5.658 & GeSn & 6.076 \\
 \hline
  InP & 5.860 & InAs & 6.050 & InSb & 6.470 & SnSi & 5.961 &   Sn & 6.489 \\
 \hline
 \hline
\end{tabular}
\label{tab:a_list}
\end{table}

\subsection{Band-gap maximization in Al$_x$Ga$_{1-x}$As}
\begin{figure}
\centering
\includegraphics[width=8.5cm]{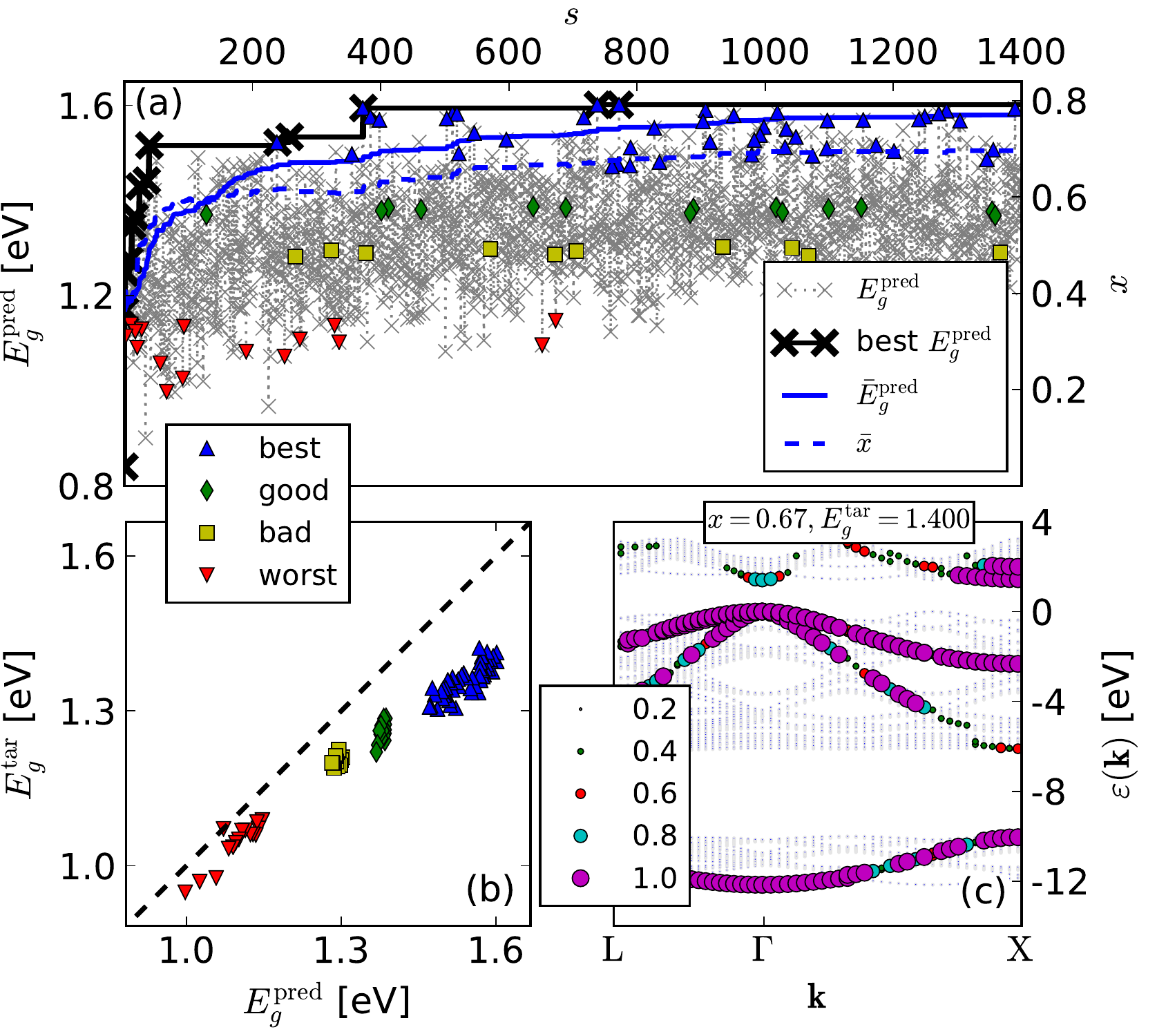}
\caption{
(a) Band gap maximization history of genetic algorithm (see Fig.~\ref{fig:genetic_algorithm}) 
based on alchemical perturbations to pure 3x3x3 GaAs super-cell. 
Predicted values, current best, average of top 20 crystals, and average mole fraction are indicated as gray crosses, black crosses, solid blue line, and dashed blue line, respectively.
Randomly selected alloys with predicted worst, bad, good, and best band-gap are respectively denoted by lower triangle, square, diamond, and circle.
(b) Scatter plot of $E_g^{\rm tar}$ vs. $E_g^{\rm pred}$ for best, good, bad, and worst alloys indicates that the overestimation of predicted band gap increases with true band gap. 
(c) Unfolded target band structure (circle size indicates weight) of converged Al$_{0.67}$Ga$_{0.33}$As alloy with largest direct gap, 
the corresponding HSE calculation yields $E_g^{\rm HSE} = 2.1$ eV.}
\label{fig:optimize}
\end{figure}

\begin{figure}
\centering
\includegraphics[width=8.5cm]{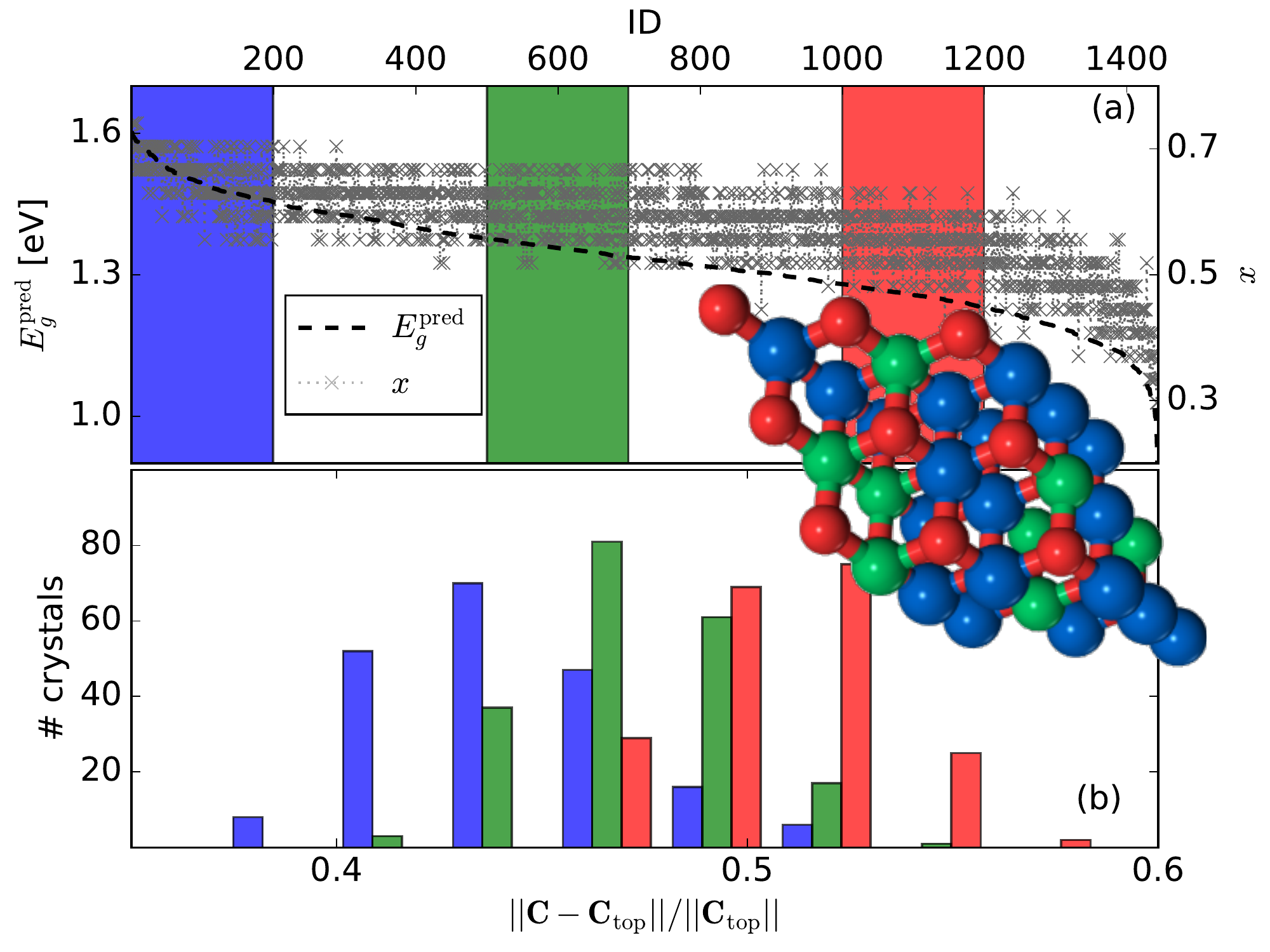}
\caption{
(a) Sorted alchemically predicted $E_g^{\rm pred}$ from optimization history in descending order (black dashed), 
superimposed with mole fraction $x$ in same order (dotted gray).
(b) Normalized sorted coulomb matrix based distance distribution between best crystal (shown as inset with Al, Ga, As in blue, green, and red, respectively) 
and crystals from three colored bands highlighted in (a) (200 crystals from each band).
}
\label{fig:population_analysis}
\end{figure}

Numerical results with such predictive power suggest that first order alchemical estimates are sufficiently accurate for robust yet efficient optimizations of Al$_x$Ga$_{1-x}$As based band structures.
Unfortunately, due to the large dimensionality of the problem, it is still challenging to optimize the band structure using reliable gradients alone. 
For example, modeling a 3x3x3 supercell of pure GaAs, corresponding to 27 Ga atoms, implies a total compositional space of 2$^{27} \sim$ 
134 million possible Al$_x$Ga$_{1-x}$As combinations (not accounting for symmetry). 
Therefore, we have used the genetic optimization algorithm, described in Fig.~\ref{fig:genetic_algorithm}
and based on alchemical perturbations towards the pseudopotential of Al at all Ga sites. 
Starting with a single unperturbed reference PBE DFT calculation, the corresponding direct band gap maximization history, 
on display in Fig.~\ref{fig:optimize}, indicates rapid convergence towards predicted PBE band gaps of 
$\sim$1.6 eV after less than 800 generations. 
The optimization history for each generation (Fig.~\ref{fig:optimize}(a)) clearly illustrates how
the average trend of direct $E_g^{\rm pred}$ is moving upward as the gradient sampling iterations proceed 
These results imply that the mating procedure during hybrid optimization \big(Fig.~\ref{fig:genetic_algorithm}(a)\big) 
systematically steers the population towards larger direct $E_g$. 
Since the band structure is determined by the structure of occupied and unoccupied orbitals,
this also indicates that crystal truncation and catenation roughly preserve the local structure of orbitals around each atom.
As a result, the algorithm identifies alloys with $E_g^{\rm pred}$ of around 1.6 eV after only several hundred iterations.
Among the best ten alloys out of 1444 identified (table~\ref{tab:opt_list}) within hybrid optimization, 
a crystal Al$_{0.67}$Ga$_{0.33}$As (structure shown in Fig.~\ref{fig:population_analysis}), with direct $E^{\rm tar}_g=1.4$ eV has been selected, and its 
unfolded\cite{band_unfold_Rubel_2014, band_unfold_zunger_2010} target band structure is plotted in Fig.~\ref{fig:optimize}(c), 
where the spectral weight at $\Gamma$ is 70.7\%, and the band gap prediction error $|\Delta E_g|=0.18$ eV.
We also note that the average value of the top 20 candidates converges quickly within $\sim$1000 generations. 


For the select examples (best, good, bad, worst) a decent correlation between $E_g^{\rm pred}$ and $E_g^{\rm tar}$ is shown in Fig.~\ref{fig:optimize}(b) 
with a linear fit yielding $\mbox{MAE}=0.014$\:eV.
Since the trends are preserved between $E_g^{\rm tar}$ and $E_g^{\rm pred}$, one genetic optimization is sufficient to identify the global optima.
Optimization performance obtained for other starting structures or larger super cells
 corroborates this observation, indicating significant robustness and generality of the design approach.
Consecutive optimizations confirm that the best alloys identified during each hybrid 
optimization are structurally similar \big(see Fig.~\ref{fig:population_analysis}(b)\big).
HSE band gaps are known to be in better agreement with experiments than PBE.~\cite{Scuseria_HSE_2004, Scuseria_HSE_2005}. 
We have therefore calculated the corresponding HSE band gaps~\cite{Scuseria_HSE_2006} 
for the ten best converged candidates, $E_g^{\rm HSE}$, listed in Tab.~\ref{tab:opt_list}.
Due to HSE being computationally more demanding, we used the PBE weights for unfolding, assuming that they will negligibly affect the HSE gap.
Throughout the optimization history, $E_g^{\rm pred}$ values cover a range of 0.9 eV to 1.6 eV, as plotted as a sorted sequence in Fig.~\ref{fig:population_analysis}(a). 
{\color{black} 
The cross-over from direct to indirect gap occurs, in line with previous calculations.~\cite{VCA, WWilkins_apl_2010, SQS_Zunger},
for mole fractions exceeding $x\approx 0.7$, substantially larger than what has been realized in experiments so far, 
i.e.~$x \approx 0.4$.}

\begin{table}
\centering
\caption{
Direct band-gaps of the ten best alchemically optimized Al$_x$Ga$_{1-x}$As alloy candidates, 
at HSE ($E_g^{\rm HSE}$), PBE ($E_g^{\rm tar}$), and alchemically ($E_g^{\rm pred}$) calculated (using the PBE band structure of pure GaAs as reference) level of theory.
Entries are ordered by $E_g^{\rm HSE}$, 
ranking for PBE (tar) and alchemical (pred) estimates are given, as well as mole fraction $x$, and spectral weight at bottom of conduction band at $\Gamma$-point ($w_\Gamma$). 
}
\resizebox{\columnwidth}{!}{%
\begin{tabular}
{c|c|c|c|c|c|R{1.5cm}} 
 \hline
 \hline
 $E_g^{\rm HSE}$ [eV] & \#(tar) & $E_g^{\rm tar}$ [eV] & \#(pred) & $E_g^{\rm pred}$ [eV] & $x$ & $w_{\Gamma}$ (\%)\\
 \hline
 \textbf{2.100} &  1 & 1.400 &  8 & 1.583 &  0.67 & 70.72 \\ 
 2.099 & 10 & \textbf{1.414} &  9 & 1.583 &  0.74 & 61.39 \\ 
 2.098 &  2 & 1.406 &  3 & 1.594 &  0.74 & 62.34 \\ 
 2.098 &  9 & 1.413 &  2 & 1.601 &  0.74 & 59.77 \\ 
 2.098 &  6 & 1.400 &  6 & 1.589 &  0.74 & 59.43 \\ 
 2.098 &  7 & 1.390 & 10 & 1.582 &  0.67 & 67.89 \\ 
 2.096 &  4 & 1.396 &  1 & \textbf{1.601} &  0.70 & 65.48 \\ 
 2.096 &  3 & 1.376 &  4 & 1.593 &  0.67 & 67.63 \\ 
 2.096 &  8 & 1.397 &  5 & 1.590 &  0.70 & 63.40 \\ 
 2.095 &  5 & 1.382 &  7 & 1.585 &  0.70 & 63.47 \\ 
 \hline
 \hline
\end{tabular}
}
\label{tab:opt_list}
\end{table}


To better understand what and how structural features impact changes in direct band gap, we have analyzed the crystal structures visited throughout the alloy optimization history. 
We compare alloys with large and small band-gap using a periodic variant of the sorted Coulomb matrix representation
$\mathbf{C}$\cite{Rupp_2012,MLcrystals_Felix2015}, where
\beq
C_{IJ} = \left\lbrace
\begin{array}{l}
0\mbox{ for } I=J\\[8pt]
\displaystyle\frac{Z_IZ_J}{\min(\fatR_{IJ})} \mbox{ else }
\end{array}
\right.
\eeq
with nuclear charges $Z$ and minimal distances $\min(\fatR_{IJ})$ between periodic images of atom positions $\fatR_I$ and $\fatR_J$.
$\mathbf{C}$ represents any crystal in a unique fashion and allows us to quantify the similarity between crystal 
$\mathbf{C}_{\rm A}$ and $\mathbf{C}_{\rm B}$ by the normalized matrix norm $\vert\vert\mathbf{C}_{\rm A} - \mathbf{C}_{\rm B}\vert\vert/\vert\vert \mathbf{C}_{\rm B}\vert\vert$.
Ranking as well as distributions of similarities to the top alloy (crystal structure shown as inset) with the widest alchemically predicted direct band gap are 
presented in Fig.~\ref{fig:population_analysis} for all alloys falling into
the 2-200 (blue), 500-700 (green), and 1000-1200 (red) windows of candidate rank. 
Overall, and not surprisingly, the top alloys clearly indicate higher probability of being more similar to the best 
candidate (blue bars larger than green than red for decreasing dissimilarity values). 
The histogram even suggests that for similarity values smaller than 0.45, 
the sorted periodic Coulomb matrix alone can be used as a descriptor to identify large direct band gap crystals with high probability.
Encouragingly, the relative mutual similarity distributions for intermediate (green) and distant (red) crystal ranks also coincide 
with the corresponding trends in band-gaps.

\begin{figure}
\centering
\includegraphics[width=8.5cm]{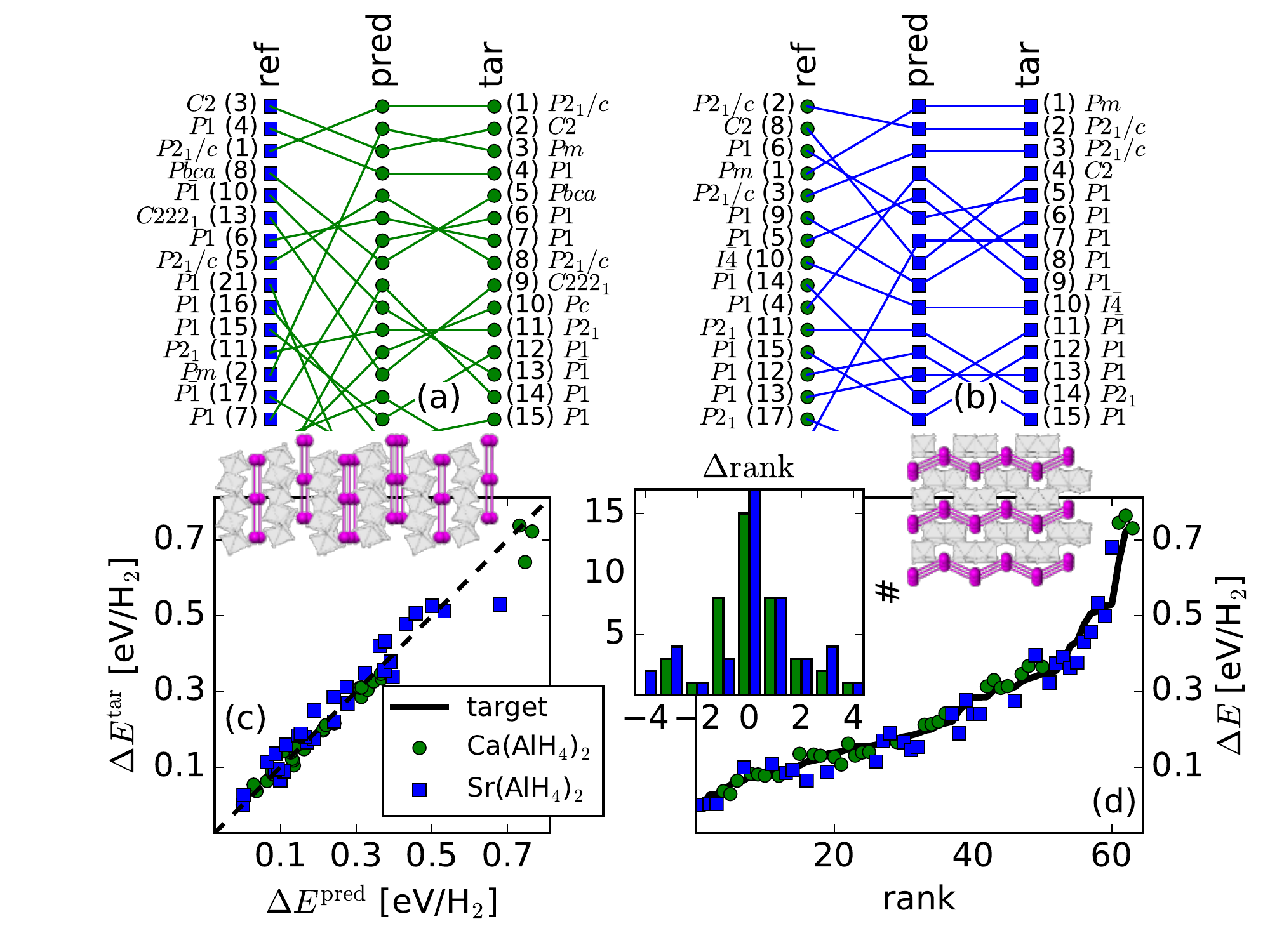}
\caption{
H$_2$ absorption energy ranking of (a) Ca-alanates (tar) predicted by first order alchemical derivatives (pred) using Sr-alanates as reference systems (ref), 
and of (b) Sr-alanates (tar) predicted by first order alchemical derivatives (pred) using Ca-alanates as reference systems (ref). 
Crystal space groups and solid lines track individual crystals.
Crystal structures of best Sr and Ca alanates are shown as an inset. 
(c) True absorption energy vs.~alchemically predicted absorption energy of Ca(Sr)-alanates in green (blue) up to 0.8 eV/H$_2$ 
(relative to the best crystal), and
(d) corresponding ranking of H$_2$ absorption energies 
by target (black line) and alchemically predicted (green and blue for predicting Ca and Sr-alanates, respectively).
Error distributions in predicted ranking ($\Delta{\rm rank}$) are shown as inset. 
}
\label{fig:order_scatter}
\end{figure}

\subsection{H$_2$ uptake in Ca and Sr alanates}
While large band-gap semi-conductors are of widespread interest, one might wonder if the alchemical approach is also applicable to other materials design problem.
In order to explore the general applicability of alchemical derivatives for ranking crystals, 
we now also consider the alchemically predicted ranking among Ca and Sr alanates with respect to their potential for H$_2$ uptake.
Ca(AlH$_4$)$_2$ and Sr(AlH$_4$)$_2$ release hydrogen when heated, and have been proposed as reversible hydrogen storage materials.\cite{alanate_nature_2001, alanate_arpcs_2009}
We have investigated 42 Ca- and 44 Sr-alanate crystal polymorph structures from Ref.~\onlinecite{Goedecker_alanate_2013}.
We comply with the aforementioned constraint of fixed structure by generating all Sr-alanate (Ca-alanate) crystals with identical crystal structures for each Ca-alanate (Sr-alanate) crystal.
The H$_2$ absorption energy of a Ca-alanate $\hat{H}^{\rm tar}$ is alchemically estimated using the Sr-alanate $\hat{H}^{\rm ref}$ in the same crystal structure.
Results in Fig.~\ref{fig:order_scatter} indicate that the alchemical estimates correctly predict the energetically strongest Ca-alanate, and the three strongest Sr-alanates.
Decent predictive power regarding the H$_2$ absorption energy ordering is also illustrated in Fig.~\ref{fig:order_scatter}(c) and (d) 
for the 60 most stable alanate crystals with a linear  
fit corresponding to a $\mbox{MAE}$ of only 0.02 eV. 
The ranking error distribution \big(Fig.~\ref{fig:order_scatter}(d) inset\big) and a Spearman's rank correlation coefficient of 0.96
also indicate substantial predictive power, sufficient for most materials design purposes.

\section{Conclusions}
To conclude and summarize, we have presented numerical evidence for the usefulness of alchemical first order derivatives 
{\color{black} for the prediction of band structures.
To illustrate the power of this method,}
we have optimized band structures in III-V semiconductors, namely  
to maximize the direct band gap in Al$_x$Ga$_{1-x}$As semiconductor alloys.
Thanks to the computational efficiency of a gradient based genetic optimization algorithm,
multiple Al$_x$Ga$_{1-x}$As alloys with direct $E_g \approx 2.1$ eV and mole fraction $x\approx 0.7$ have been identified. 
{\color{black} We note that after having identified such promising materials candidates, 
the synthetic procedure to realize them in an experiment remains a largely outstanding challenge.}
The qualitative identification of alloys with the largest $E_g$ within single optimization runs implies that the $E_g$ surface 
is fairly flat in the crystal space of Al$_x$Ga$_{1-x}$As systems.
{\color{black} 
Alchemical derivatives can also tackle other challenges in the realm of computational materials design, 
such as estimating H$_2$ absorption energy ordering across Ca and Sr-alanates of varying crystal structure. 
}
Future work will deal with {\color{black} (i) extensions to higher orders, and (ii) systematic
comparisons to alternative materials design approaches such as special quasirandom structures~\cite{SQS},
cluster expansion~\cite{Clusterexpansion}, or machine learning approaches~\cite{Faber_prl_2016}.}

\section*{Acknowledgements}
OAvL acknowledges support by the Swiss National Science foundation (No.~PP00P2\_138932, 407540\_167186 NFP 75 Big Data, 200021\_175747, NCCR MARVEL).
Some calculations were performed at sciCORE (http://scicore.unibas.ch/) scientific computing core facility at University of Basel.

\appendix*
\section{Gradient based genetic optimization}

\begin{figure}
\centering
\includegraphics[width=8.5cm]{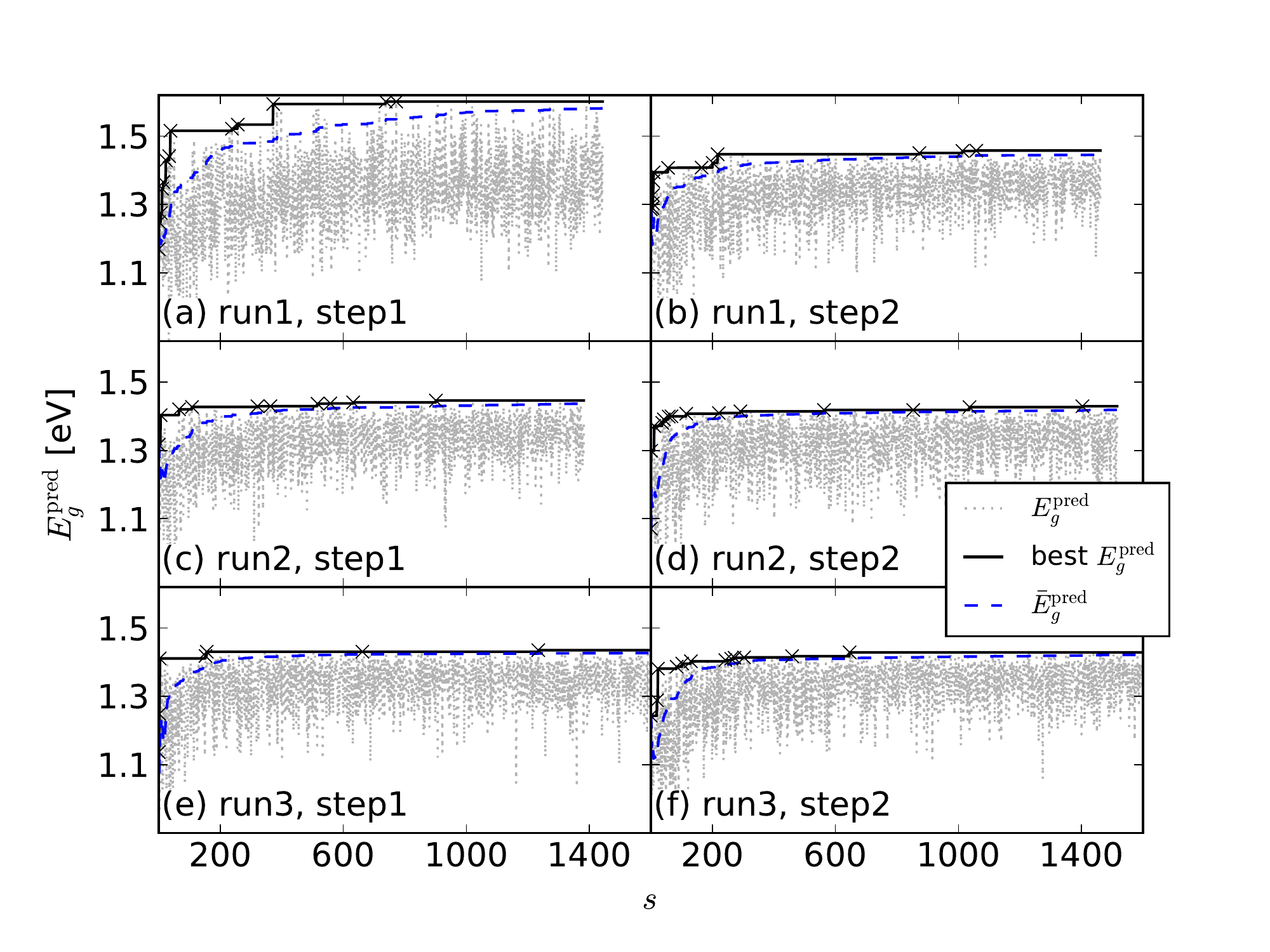}
\includegraphics[width=8.5cm]{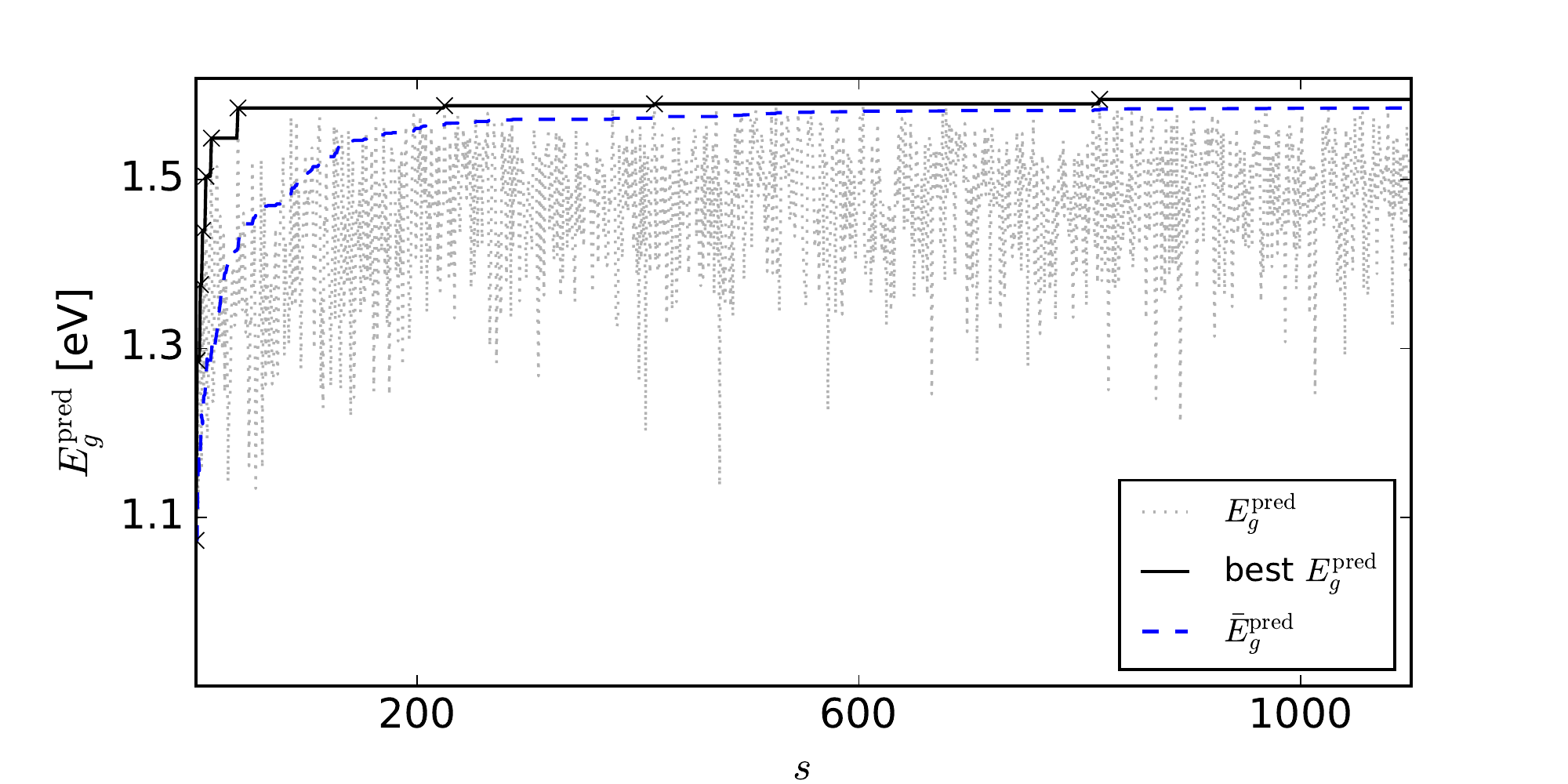}
\caption{
LEFT: Three optimization runs with different initial distributions of atoms, once pristine GaAs (top), and twice Ga/Al distributed at random (left), 
and after restarting the optimization using the wave function of the converged system (right) in 3$\times$3$\times$3 super cells.
RIGHT: Optimization run starting with random initial atomic positions in 4$\times$4$\times$4 super cell.
}
\label{fig:optimization}
\end{figure}

The objective of the gradient-based genetic algorithm (GA) is to maximize the band gap while the bottom of the conduction band must be located at $\Gamma$-point in the Brillouin zone.
To this end, we vary the mixing ratio of Al and Ga, as well as their location.
Instead of performing a DFT evaluation for each candidate crystal, 
the band structure of a candidate crystal is estimated using alchemical derivatives.
Each alchemical prediction takes less than 1\% of the computational cost for a full DFT band structure calculation.
This enables the rapid exploration of many configurations and mixture ratios with decent accuracy.
The Brillouin zone of the supercell is unfolded and corresponds to the one from a primitive cell using spectral weights.\cite{band_unfold_Rubel_2014, band_unfold_zunger_2010}

Three independent gradient-based GA optimizations have been carried out,
start with three starting crystals: Pure GaAs and two random crystals, where either Ga or Al is randomly chosen for each of 27 III-sites. 
All three optimizations converged within two optimization steps, which give total six GA sampling histories.
3137 crystals have been evaluated in the first GA sampling history.
Out of these, 1444 are predicted to be direct band gap crystal.
Roughly half of the searched crystals are of indirect band gap. 
$E_g^{\rm true}$ are calculated for the following four groups of crystals from the first GA sampling step history,
depending on the sorted direct band gap predictions: 
the best 50 $E_g^{\rm best}$,
the top 491 to 504 $E_g^{\rm good}$,
the top 1001 to 1010 $E_g^{\rm bad}$,
and the worst 15 $E_g^{\rm worst}$.
The analysis of the first GA sampling history is shown in Fig.~\ref{fig:optimize}, while the analysis of all three optimization histories is presented in Fig.~\ref{fig:population_analysis}.
Pseudocode for the gradient-based optimization is presented in the pseudo-code below, where each of the routines corresponds to the following.
\begin{itemize}
\item \textbf{optimization}: The main routine consist of: A stochastic sampling procedure; and a gradient step that updates wavefunction by full DFT evaluation, 
as schematically shown in the left panel of Fig.~\ref{fig:genetic_algorithm}.
The convergence criterion $|E_g^{h} - E_g^{(h-1)}| < 0.001$ eV at $h^{th}$ optimization step.
\item \textbf{full DFT}: It performs full DFT evaluations on the requested crystals and returns the best corresponding true band gap, $E_g^{\rm true}$, and orbitals, $\{\phi_i\}$. 
\item \textbf{GA sampling}: The routine stochastically samples the alchemical estimates, using reference orbitals $\{\phi_i\}$. 
The initial population of twenty crystals is randomly generated.
The best five crystals are used for full DFT evaluation. 
The sampling criterion is at least 1400 crystals with direct band gap are found. And the difference between the average of the top-20 population, $\bar{E}_g^{\rm pred}$, and the best-predicted crystal, $E_g^{\rm best}$, is less than 0.02 eV.
\item \textbf{get parents}: At each iteration during GA sampling, two parents, {\tt ParentA} and {\tt ParentB}, are randomly drawn from the top-20 of all searched crystals. 
In other words, the population size of GA is twenty.
\item \textbf{mate}: The routine to generate a child from two parent crystals.
The occupancy at each of the III-sites in the child crystal is randomly inherited from either parent with 50/50 possibility, as illustrated in the right panel of Fig.~\ref{fig:genetic_algorithm}.
A mutation rate of 0.05\% is used at each of the III-sites, where the atom type would flip from Al to Ga or vise versa as indicated the yellow atom in the right panel of Fig.~\ref{fig:genetic_algorithm}.
\item \textbf{alchemical evaluation} estimate the band structure of the requested crystal using reference orbital $\{\phi_i\}$. 
If the prediction of the band gap is indirect, the value of the direct band gap is set to 0.
\end{itemize}

\begin{algorithm}
\caption{
Gradient-based/genetic algorithm using alchemical derivative for crystal optimization. The pseudocode of the main routine, 
optimization, and two primary functions, full DFT and GA sampling, are explicitly stated. The variables are denoted by {\tt typewriter} font.
}
\begin{algorithmic}[100]
\Procedure{\textbf{optimization}}{{\tt starting\_crystal}}
\State $E_g^{(0)}\gets 0$
\State $h\gets 1$
\State $(E_g^{(h)}, \{\phi_i^{(h)}\}) \gets$ \textbf{full DFT}({\tt starting\_crystal})
\While {$|E_g^{(h)} - E_g^{(h-1)}| > 0.001\:$eV}
\State {\tt top\_5\_predicted\_crystals} $\gets$ \textbf{GA sampling}($\{\phi_i^{(h)}\}$)
\State $(E_g^{\rm (h+1)}, \{\phi_i^{(h+1)}\})$ $\gets$ \textbf{full DFT}({\tt top\_5\_predicted\_crystals})
\State $h\gets h+1$
\EndWhile
\EndProcedure
\\
\Function{\textbf{full DFT}}{{\tt crystals}}
\State preform full DFT evaluations on every {\tt crystals}
\State \Return $(E_g^{\rm true}, \{\phi_i\})$ of the best crystal among {\tt crystals}
\EndFunction
\\
\Function{\textbf{GA sampling}}{$\{\phi_i\}$}
\State initialize {\tt population} of size 20
\State $E_g^{\rm pred} \gets$ \textbf{alchemical evaluation}($\{\phi_i\}$, {\tt population})
\State $s\gets 1$
\While {$s < 1400$ \textbf{and} $|\bar{E}_g^{\rm pred} - E_g^{\rm best}| < 0.02$ eV}
\State ({\tt ParentA, ParentB}) $\gets$ \textbf{get parent}({\tt population})
\State {\tt Child} $\gets$ \textbf{mate}({\tt ParentA, ParentB})
\If {{\tt Child} is not in {\tt population}}
\State $E_g^{\rm pred} \gets$ \textbf{alchemical evaluation}($\{\phi_i\}$, {\tt Child})
\State update {\tt population}
\If {$E_g^{\rm pred}$ is direct}
\State $s\gets s+1$
\EndIf
\EndIf
\EndWhile
\State \Return top 5 crystals in the {\tt population}
\EndFunction
\end{algorithmic}
\label{alg:gradient_GA}
\end{algorithm}

\begin{figure}
\centering
\includegraphics[width=8.5cm]{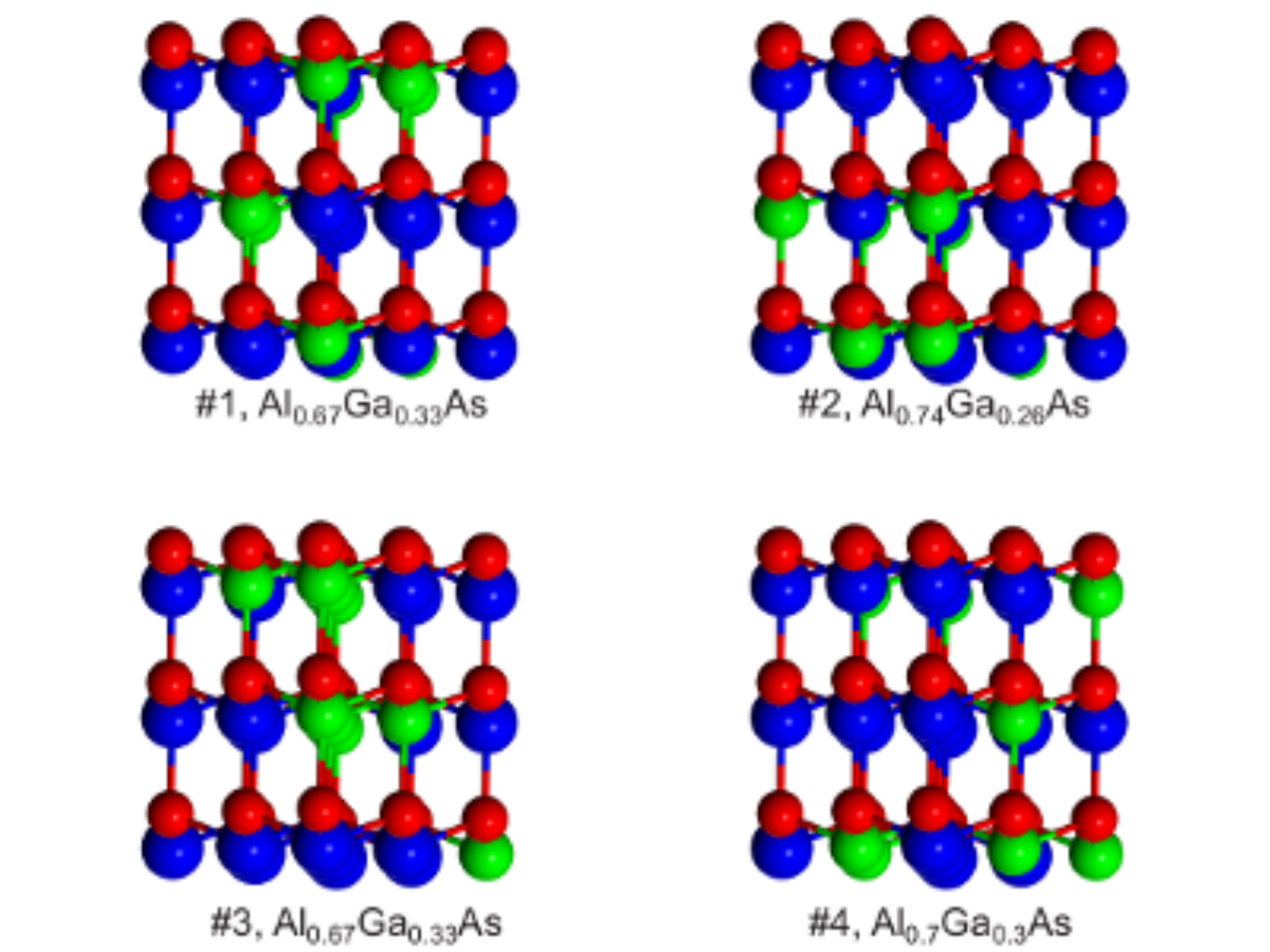}
\caption{The best four crystals with $E_g^{\rm HSE}\approx 2.1$ eV according to Table.~\ref{tab:opt_list}.
The HSE ordering rank and the crystal formula with mole fraction is shown for each crystal. 
The $3\times 3\times 3$ supercell is extended to $6\times 6\times 6$ to illustrate 2-fold symmetry, where Al, Ga, and As are represented by blue, green, and red spheres.
}
\label{fig:top_crystals}
\end{figure}

\begin{table}
\caption{3$\times$3$\times$3 super-cells for 10 top crystals with large direct band-gap (in the order of predicted gap in Table I in main text). 
Each column corresponds to a crystal, each row to the atomic site.
Upper and lower block correspond to III and V elements, respectively}
\begin{tabular}{l|c|c|c|c|c|c|c|c|c|c|c}
\hline\hline
rank & 1 & 2 & 3 & 4&5&6&7&8&9&10&$(x,y,z)$ [\AA]\\
\hline
 1 & Al & Ga & Al & Ga & Ga & Al & Al & Ga & Al & Ga & ( 0.0,  0.0,  0.0) \\
 2 & Ga & Al & Ga & Al & Ga & Al & Al & Al & Ga & Al & ( 4.0,  0.0,  0.0) \\
 3 & Al & Al & Al & Ga & Al & Al & Al & Ga & Al & Al & ( 8.0,  0.0,  0.0) \\
 4 & Al & Al & Al & Ga & Al & Ga & Al & Al & Al & Al & ( 2.0,  3.5,  0.0) \\
 5 & Ga & Ga & Al & Ga & Al & Ga & Al & Ga & Ga & Ga & ( 6.0,  3.5,  0.0) \\
 6 & Al & Ga & Al & Al & Al & Al & Ga & Al & Al & Al & (10.0,  3.5,  0.0) \\
 7 & Al & Al & Al & Al & Al & Al & Al & Al & Al & Al & ( 4.0,  6.9,  0.0) \\
 8 & Al & Al & Ga & Al & Ga & Ga & Ga & Al & Ga & Ga & ( 8.0,  6.9,  0.0) \\
 9 & Al & Al & Al & Al & Al & Ga & Al & Al & Al & Ga & (12.0,  6.9,  0.0) \\
10 & Al & Al & Al & Al & Al & Ga & Al & Ga & Ga & Al & ( 2.0,  1.2,  3.3) \\
11 & Al & Al & Al & Al & Ga & Al & Al & Al & Al & Al & ( 6.0,  1.2,  3.3) \\
12 & Al & Ga & Ga & Al & Al & Al & Ga & Ga & Al & Al & (10.0,  1.2,  3.3) \\
13 & Ga & Al & Al & Ga & Al & Al & Ga & Al & Al & Ga & ( 4.0,  4.6,  3.3) \\
14 & Al & Al & Al & Al & Ga & Al & Al & Al & Al & Al & ( 8.0,  4.6,  3.3) \\
15 & Al & Al & Ga & Al & Al & Al & Al & Al & Al & Al & (12.0,  4.6,  3.3) \\
16 & Al & Al & Al & Al & Al & Al & Ga & Al & Ga & Ga & ( 6.0,  8.1,  3.3) \\
17 & Ga & Al & Ga & Al & Al & Ga & Al & Ga & Al & Al & (10.0,  8.1,  3.3) \\
18 & Al & Al & Al & Ga & Al & Al & Ga & Al & Al & Ga & (14.0,  8.1,  3.3) \\
19 & Al & Al & Al & Ga & Al & Al & Al & Al & Al & Al & ( 4.0,  2.3,  6.5) \\
20 & Ga & Ga & Al & Al & Al & Al & Ga & Al & Al & Ga & ( 8.0,  2.3,  6.5) \\
21 & Al & Al & Ga & Al & Al & Al & Al & Al & Al & Al & (12.0,  2.3,  6.5) \\
22 & Al & Al & Al & Al & Al & Al & Al & Al & Al & Al & ( 6.0,  5.8,  6.5) \\
23 & Al & Ga & Al & Ga & Al & Ga & Al & Al & Al & Ga & (10.0,  5.8,  6.5) \\
24 & Ga & Ga & Al & Al & Al & Al & Al & Ga & Al & Al & (14.0,  5.8,  6.5) \\
25 & Ga & Al & Al & Al & Ga & Al & Al & Ga & Ga & Al & ( 8.0,  9.2,  6.5) \\
26 & Ga & Al & Al & Al & Ga & Al & Ga & Al & Ga & Al & (12.0,  9.2,  6.5) \\
27 & Al & Al & Ga & Ga & Ga & Al & Al & Ga & Al & Al & (16.0,  9.2,  6.5) \\
\hline
28 & As & As & As & As & As & As & As & As & As & As & ( 2.0,  1.2,  0.8) \\
29 & As & As & As & As & As & As & As & As & As & As & ( 6.0,  1.2,  0.8) \\
30 & As & As & As & As & As & As & As & As & As & As & (10.0,  1.2,  0.8) \\
31 & As & As & As & As & As & As & As & As & As & As & ( 4.0,  4.6,  0.8) \\
32 & As & As & As & As & As & As & As & As & As & As & ( 8.0,  4.6,  0.8) \\
33 & As & As & As & As & As & As & As & As & As & As & (12.0,  4.6,  0.8) \\
34 & As & As & As & As & As & As & As & As & As & As & ( 6.0,  8.1,  0.8) \\
35 & As & As & As & As & As & As & As & As & As & As & (10.0,  8.1,  0.8) \\
36 & As & As & As & As & As & As & As & As & As & As & (14.0,  8.1,  0.8) \\
37 & As & As & As & As & As & As & As & As & As & As & ( 4.0,  2.3,  4.1) \\
38 & As & As & As & As & As & As & As & As & As & As & ( 8.0,  2.3,  4.1) \\
39 & As & As & As & As & As & As & As & As & As & As & (12.0,  2.3,  4.1) \\
40 & As & As & As & As & As & As & As & As & As & As & ( 6.0,  5.8,  4.1) \\
41 & As & As & As & As & As & As & As & As & As & As & (10.0,  5.8,  4.1) \\
42 & As & As & As & As & As & As & As & As & As & As & (14.0,  5.8,  4.1) \\
43 & As & As & As & As & As & As & As & As & As & As & ( 8.0,  9.2,  4.1) \\
44 & As & As & As & As & As & As & As & As & As & As & (12.0,  9.2,  4.1) \\
45 & As & As & As & As & As & As & As & As & As & As & (16.0,  9.2,  4.1) \\
46 & As & As & As & As & As & As & As & As & As & As & ( 6.0,  3.5,  7.3) \\
47 & As & As & As & As & As & As & As & As & As & As & (10.0,  3.5,  7.3) \\
48 & As & As & As & As & As & As & As & As & As & As & (14.0,  3.5,  7.3) \\
49 & As & As & As & As & As & As & As & As & As & As & ( 8.0,  6.9,  7.3) \\
50 & As & As & As & As & As & As & As & As & As & As & (12.0,  6.9,  7.3) \\
51 & As & As & As & As & As & As & As & As & As & As & (16.0,  6.9,  7.3) \\
52 & As & As & As & As & As & As & As & As & As & As & (10.0, 10.4,  7.3) \\
53 & As & As & As & As & As & As & As & As & As & As & (14.0, 10.4,  7.3) \\
54 & As & As & As & As & As & As & As & As & As & As & (18.0, 10.4,  7.3) \\
\hline\hline
\end{tabular}
\end{table}

\bibliography{literature}{}
\bibliographystyle{ieeetr}
\end{document}